\documentclass[onecolumn]{aa}
\usepackage{graphicx}
\usepackage{txfonts}

\def\mean#1{{\langle}#1{\rangle}}

\begin{document}
   \title{Constraints to Uranus' Great Collision IV}

   \subtitle{The Origin of Prospero}

   \author{M.  Gabriela  Parisi   \inst{1,2,3}\fnmsep\thanks {Member  of  the
          Carrera  del  Investigador   Cient\'{\i}fico,  Consejo  Nacional  de
          Investigaciones  Cient\'{\i}ficas y T\'ecnicas  (CONICET), Argentina}
          \and
          Giovanni Carraro\inst{4,5}  \and Michele Maris  \inst{6} \and Adrian
          Brunini \inst{3} } \offprints{M. Gabriela Parisi}

         \institute{Instituto Argentino de Radioastronom\'{\i}a (IAR),
             C.C. N$^{o}$ 5, 1894 Villa Elisa, Argentina\\
             \email{gparisi@iar-conicet.gov.ar}
         \and
             Departamento de Astronom\'{\i}a, Universidad de Chile,
              Casilla 36-D, Santiago, Chile\\
              \email{gparisi@das.uchile.cl}
         \and
             Facultad de Ciencias Astron\'omicas y Geof\'{\i}sicas,
             Universidad Nacional de La Plata, Argentina\\
             \email{gparisi,abrunini@fcaglp.fcaglp.unlp.edu.ar}           
          \and
              European Southern Observatory (ESO), Alonso de Cordova 3107,
              Vitacura, Santiago, Chile\\ 
              \email{gcarraro@eso.org}
          \and
              Dipartamento di Atronomia, Universit\'a di Padova,
              Vicolo Osservatorio 2, I-35122 Padova, Italy\\ 
              \email{giovanni.carraro@unipd.it}         
          \and
             INAF, Osservatorio Astronomico di Trieste, Via G.B. 
             Tiepolo 11, I-34131 Trieste, Italy\\
             \email{maris@oats.inaf.it}             }

   \date{Received June 2007 ; accepted  January 2008}

\abstract{ 
%$Context.$ 
It  is widely accepted  that the large  obliquity of Uranus  is the
result of a great tangential collision (GC) with an Earth size proto-planet at
the end of the accretion process.  The impulse imparted by the GC had affected
the  Uranian  satellite  system.   Very recently,  nine  irregular  satellites
(irregulars) have  been discovered around Uranus.  Their  orbital and physical
properties, in particular those of  the irregular Prospero, set constraints on
the  GC scenario.  
}{
%$Aims.$
We attempt to  set constraints  on the  GC scenario  as the  cause of
Uranus' obliquity  as well  as on the  mechanisms able  to give origin  to the
Uranian irregulars.
}{
%$Methods.$
Different  capture mechanisms  for irregulars operate  at different
stages on the giant planets  formation process. The mechanisms able to capture
the uranian  irregulars before and after  the GC are  analysed.  Assuming that
they were  captured before the  GC, we calculate  the orbital transfer  of the
nine irregulars by the impulse imparted  by the GC.  If their orbital transfer
results dynamically implausible, they should have originated after the GC.  We
then investigate and discuss the  dissipative mechanisms able to operate later.
}{
%$Results.$ 
Very  few transfers exist for  five of the  irregulars, which makes
their existence before  the GC hardly expected.  In  particular Prospero could
not exist  at the time of  the GC.  Different capture  mechanisms for Prospero
after  the GC  are investigated.   Gas drag  by Uranus'envelope  and pull-down
capture  are  not  plausible   mechanisms.   Capture  of  Prospero  through  a
collisionless interaction  seems to  be difficult.  The  GC itself  provides a
mechanism of permanent capture.  However, the capture of Prospero by the GC is
a low probable  event.  Catastrophic collisions could be  a possible mechanism
for the birth  of Prospero and the other irregulars after  the GC. Orbital and
physical clusterings should then be expected.
}{
%$Conclusions$
Either Prospero had to originate  after the GC or the GC did not
occur.  In the former case, the mechanism for the origin of Prospero after the
GC remains an  open question. An observing program able  to look for dynamical
and physical  families is  mandatory.  In the  latter case, another  theory to
account  for  Uranus' obliquity  and  the  formation  of the  Uranian  regular
satellites on the equatorial plane of the planet would be needed.

\keywords{Planets and satellites: general --
                Planets and satellites: formation--
                Solar System: general--
                   Solar System: formation
               }
   }

   \maketitle

%
%--------------------------------------------------------------------------%

\section{Introduction}

Very recently, rich systems  of irregular satellites (hereafter irregulars) of
the giant  planets have been discovered.   Enabled by the  use of large-format
digital  images  on  ground-based  telescopes,  new  observational  data  have
increased the  known population  of Jovian irregulars  to 55 (Sheppard  et al.
\cite{shepparda}),   the  Saturnian   population   to  38   (Gladman  et   al.
\cite{gladmanb}, Sheppard et  al.  \cite{sheppardb}, \cite{shepparde}) and the
Neptunian  population to  7 (Holman  et  al.  \cite{holman},  Sheppard et  al.
\cite{sheppardd}).   The Uranian  system  is of  particular  interest since  a
population  of  9  irregulars  (named  Caliban,  Sycorax,  Prospero,  Setebos,
Stephano,  Trinculo, S/2001U2:  XXIV Ferdinand,  S/2001U3: XXII  Francisco and
S/2003U3: XXIII  Margaret) has been  discovered around Uranus (Gladman  et al.
\cite{gladman}, \cite{gladmana}, Kavelaars  et al.  \cite{kavelaars}, Sheppard
et al.  \cite{sheppardc}).   The discovery of these objects  provides a unique
window on  processes operating  in the young  Solar System. In  the particular
case of  Uranus, their existence may  cast light on  the mechanism responsible
for its  peculiar rotation axis  (Parisi \& Brunini \cite{parisi},  Brunini et
al.  \cite{brunini} (hereafter BP02)).

Irregulars of giant planets are characterized by eccentric, highly tilted with
respect of  the parent planet equatorial  plane, and in  some case retrograde,
orbits.  These objects cannot have  formed by circumplanetary accretion as the
regular  satellites  but they  are  likely products  of  an  early capture  of
primordial  objects from  heliocentric  orbits, probably  in association  with
planet formation  itself (Jewitt \& Sheppard \cite{jewitta}).   It is possible
for an object circling about the sun to be temporarily trapped by a planet. In
terms of the classical three-body problem  this type of capture can occur when
the object passes through the  interior Lagrangian point, $L_{2}$, with a very
low relative  velocity.  But, without any  other mechanism, such  a capture is
not permanent  and the objects will  eventually return to a  solar orbit after
several or several hundred orbital  periods.  To turn a temporary capture into
a  permanent one  requires a  source of  orbital energy  dissipation  and that
particles could remain  inside the Hill sphere long enough  for the capture to
be effective.

Although  currently  giant  planets  have  no efficient  mechanism  of  energy
dissipation for permanent capture, at their formation epoch several mechanisms
may have operated:  1) {\em gas drag}  in the solar nebula or  in an extended,
primordial  planetary atmosphere  or  in a  circumplanetary  disk (Pollack  et
al.\cite{pollack}, Cuk \& Burns \cite{cuk}), 2) {\em pull-down capture} caused
by the  mass growth and/or orbital  expansion of the planet  which expands its
Hill    sphere    (Brunini     \cite{bruninia},    Heppenheimer    \&    Porco
\cite{heppenheimer}),  3) {\em collisionless  interactions} between  a massive
planetary satellite and guest bodies  (Tsui \cite{tsui}) or between the planet
and  a binary  object (Agnor  \&  Hamilton \cite{agnor}),  and 4)  collisional
interaction between  two planetesimals  passing near the  planet or  between a
planetesimal and a regular satellite.  This last mechanism, the so called {\em
break-up} process, leads to the formation of dynamical groupings (e.g. Colombo
\& Franklin  1971, Nesvorny  et al. \cite{nesvornya}).   After a  break-up the
resulting fragments of  each progenitor would form a  population of irregulars
with similar surface composition,  i.e.  similar colors, and irregular shapes,
i.e.   light-curves  of  wide  amplitude.   Significant  fluctuations  in  the
light-curves of  Caliban (Maris et  al.  \cite{maris}) and Prospero  (Maris et
al.   \cite{marisa}) and  the  time  dependence observed  in  the spectrum  of
Sycorax (Romon  et al.  \cite{romon}) suggest  the idea of  a break-up process
for the origin of these bodies.

Several  theories to  account  for the  large  obliquity of  Uranus have  been
proposed.   Kubo-Oka  \& Nakazawa  (\cite{kubo-oka}),  investigated the  tidal
evolution of  satellite orbits and  examined the possibility that  the orbital
decay of  a retrograde satellite leads  to the large obliquity  of Uranus, but
the large mass required for  the hypothetical satellite makes this possibility
very  implausible.    An  asymmetric  infall  or  torques   from  nearby  mass
concentrations during the collapse of  the molecular cloud core leading to the
formation of the  Solar System, could twist the  total angular momentum vector
of the  planetary system.   This twist could  generate the obliquities  of the
outer  planets (Tremaine  \cite{tremaine}). This  model has  the disadvantages
that the  outer planets must form before  the infall is complete  and that the
conditions for the event that would  produce the twist are rather strict.  The
model  itself is  difficult  to  be quantitatively  tested.   Tsiganis et  al.
(\cite{tsiganis}) proposed that the  current orbital architecture of the outer
Solar System could have been  produced from an initially compact configuration
with  Jupiter and  Saturn  crossing  the 2:1  orbital  resonance by  divergent
migration.   The crossing  led to  close encounters  among the  giant planets,
producing   large   orbital  eccentricities   and   inclinations  which   were
subsequently  damped to the current value by gravitational interactions with
planetesimals.   The  obliquity changes  due  to  the  change in  the  orbital
inclinations.  Since the inclinations are damped by planetesimals interactions
on  timescales much  shorter than  the timescales  for precession  due  to the
torques from the Sun, especially for Uranus and Neptune, the obliquity returns
to small values if it is small before the encounters (Hoi et al. \cite{hoi}).

Large stochastic impacts at the  last stage of the planetary formation process
have  been  proposed  as  the  possible cause  of  the  planetary  obliquities
(e.g. Safronov  \cite{safronov}).  The large obliquity  of Uranus (98$^\circ$)
is usually attributed to a  great tangential collision (GC) between the planet
and an Earth-size  planetesimal occurred at the end of  the epoch of accretion
(e.g., Parisi \& Brunini \cite{parisi}, Korycansky et al.  \cite{korycansky}).
The collision imparts  an impulse to Uranus and  allows preexisting satellites
of the  planet to change  their orbits.  Irregulars  on orbits with  too large
semimajor  axis  escape  from  the  system (Parisi  \&  Brunini  1997),  while
irregulars  with a  smaller semimajor  axis may  be pushed  to outer  or inner
orbits  acquiring greater  or lower  eccentricities depending  on  the initial
orbital elements, the geometry of the impact and the satellite position at the
moment of impact.  The orbits  excited by this perturbation must be consistent
with  the present  orbital  configuration of  the  Uranian irregulars  (BP02).

In an attempt to clarify the origin of Uranus obliquity and of its irregulars,
we are using  in this study the most updated information  on their orbital and
physical properties.

 In Section  2, we improve  the model developed  in BP02 for the  five Uranian
 irregulars known at  that epoch and extend our study to  the new four Uranian
 irregulars recently  discovered by  Kavelaars et al.   (\cite{kavelaars}) and
 Sheppard et al. (\cite{sheppardc}).  The origin of these objects after the GC
 is  discussed  in Section  3,  where several  mechanisms  for  the origin  of
 Prospero are investigated.  The discussion of the results and the conclusions
 are presented in Section 4.

%--------------------------------------------------------------------------%

\section{Transfer of the irregulars to their current  orbits:}

Assuming the GC scenario, the transfers  of the nine known irregulars to their
current orbits are computed following  the procedure developed in BP02 for the
five irregulars known in 2002.   We present improved calculations using a more
realistic  code to  compute the  evolution of  the irregulars  current orbital
eccentricities.

If the  large obliquity of  Uranus has been  the result of a  giant tangential
impact,  the orbits  of  preexisting  satellites changed  due  to the  impulse
imparted to  the planet  by the collision.   The angular momentum  and impulse
transfer to the Uranian system at impact were modeled using the Uranus present
day rotational and orbital properties as imput parameters (BP02).

Just  before the  GC, the square  of the  orbital velocity $\nu_{1}$  of a
preexisting satellite of negligible mass is given by:

\begin{equation}  \nu_{1}^{2}  =  G  m_{U}  \bigg( {{2}  \over{r}}-  {{1}
\over{a_{1}}} \bigg), \label{eq:v1} \end{equation}

\noindent $r$ being  the position of the satellite on its  orbit at the moment
of the  GC, $a_{1}$ its orbital semiaxis  and $m_U$ the mass  of Uranus before
the impact. The impactor mass is  $m_i$ and $G$ is the gravitational constant.
After  the GC, the  satellite is  transferred to  another orbit  with semiaxis
$a_{2}$ acquiring the following square of the velocity:

\begin{equation}  \nu_{2}^{2}  =  G  (m_{U}+m_{i}) \bigg( {{2} \over{r}}-
{{1} \over{a_{2}}} \bigg). \label{eq:v2} \end{equation}

We set $\nu_{1}^{2}= A\; \nu_{e}^{2}$  and $\nu_{2}^{2}= B\; ( 1+ m_{i}/m_{U})
\; \nu_{e}^{2}$, where $A$ and $B$ are arbitrary coefficients ($0 < A \leq 1$,
$ B >0$), $\nu_{e}$ being the escape velocity at $r$ before the GC.

The  semiaxis of the  satellite orbit  before ($a_1$)  and after  ($a_2$) the
GC verify the following simple relations:

\begin{equation}
a_1=\frac{r}{2\;(1-A)} \;\;,\;\;\; a_2=\frac{r}{2\;(1-B)}\;.
\label{a1a2}
\end{equation}

If $A < B$ then $a_{1} < a_{2}$.  In the special case of $B=1$, the orbits are
unbound from the system. If $A > B$ then $a_{1} > a_{2}$, the initial orbit is
transferred  to an  inner orbit.   When  $A=B$, the  orbital semiaxis  remains
unchanged ($a_{1} = a_{2}$).

The position $r$ of the satellite on  its orbit at the epoch of the impact may
be expressed in the following form:

\begin{equation}
r=\frac{2\; G\; m_U}{(\Delta V)2} \
\bigg[ \frac{B'-A} { \sqrt{A} \cos{\Psi} \pm
\sqrt{ (B'-A) + A \cos2{\Psi} }
} \bigg]2,
\label{r}
\end{equation} 

\noindent with  $B'= B\; (  1+ m_{i}/m_{U})$.  Since stochastic  processes can
only take  place at  very late  stages in the  history of  planetary accretion
(e.g. Lissauer  \& Safronov \cite{lissauera}), the  GC is assumed  to occur at
the end of Uranus formation (e.g.  Korycansky et al.  \cite{korycansky}).  The
mass of  Uranus after the GC,  ($m_i+m_U$), is taken as  Uranus' present mass.
$\Psi$  is the  angle between  $\vec{\nu_1}$ and  the orbital  velocity change
imparted   to  Uranus   $\Delta{\vec{V}}$.   An   analytical   expression  for
$\Delta{\vec{V}}$ is  derived in  BP02 assuming that  the impact  is inelastic
(Korycansky  et al.   \cite{korycansky}) as  a function  of $m_i$,  the impact
parameter  of the  collision $b$,  the  present rotation  angular velocity  of
Uranus $\vec{\Omega}$, the spin angular velocity which Uranus would have today
if the collision had not  occurred $\vec{\Omega}_0$, and $\alpha$ which is the
angle between $\vec{\Omega}$ and $\vec{\Omega}_0$:

\begin{equation}   \Delta{\vec{V}}  =   {{2 R_{U}^{2}}\over {5  b}}
\bigg[    \Omega^{2}+     {
{\Omega_{0}^{2}}\over{              (1+{{m_{i}}\over{m_{U}}})^{2}
(1+{{m_{i}}\over{3m_{U}}})^{4}}}   -   {   {2  \Omega  \Omega_{0}
\cos{\alpha}}\over{
(1+{{m_{i}}\over{m_{U}}})(1+{{m_{i}}\over{3m_{U}}})^{2}}}\bigg]^{1/2},
\label{vi} \end{equation}

\noindent $R_U$  being the present  equatorial radius of Uranus.   A collision
with the core itself was necessary  in order to impart the required additional
mass and  angular momentum (Korycansky et al.   \cite{korycansky}).  Since $b$
is an unknown quantity, we take its most probable value: $b$=(2/3)$R_C$, where
$R_C$ is the  core radius of Uranus  at the moment of collision  assumed to be
1.8 $\times$ 10$^{4}$ km (Korycansky et al.  \cite{korycansky}, Bodenheimer \&
Pollack \cite{bodenheimer}).   The results have  a smooth dependence  with the
impactor mass  which allows us to  take $m_i$ $\sim$  1$m_{\oplus}$ (Parisi \&
Brunini \cite{parisi}).

The  minimum eccentricity of the orbits before the collision is given
by:

\begin{eqnarray}
e_{1min}= 2 \;( 1 - A ) - 1 \;\;\;\; if\;\;\;\; A \leq 0.5\;\;\;,\;\;
e_{1min}= 1 - 2 \;( 1 - A ) \;\;\;\; if\;\;\;\; A > 0.5,
\label{e1m}
\end{eqnarray}

\noindent while the minimum eccentricity of the orbits after the collision is:

\begin{eqnarray}
e_{2min}= 2 \;( 1 - B )- 1 \;\;\;\; if\;\;\;\; B \leq 0.5\;\;\;,\;\;
e_{2min}= 1 - 2 \;( 1 - B ) \;\;\;\; if\;\;\;\; B > 0.5. 
\label{e2m}
\end{eqnarray}

The minimum possible value of $\Delta{V}$ ($\Delta{V}_{min}$) is obtained from
Eq.~(\ref{vi}) for an initial period  $T_0$= 20 hrs ($T_0= 2\pi/\Omega_0$) and
$\alpha=70^{o}$  (BP02). Therefore,  although simulations  of  solid accretion
produce in general random  spin orientations (e.g.  Chambers \cite{chambers}),
we  further assume  $T_0$= 20  hrs  and $\alpha$=70$^{o}$  in order  to set  a
maximum bound  on Eq.~(\ref{r}).  Upper  bounds in $a_1$ ($a_{1M}$)  and $a_2$
($a_{2M}$)  are  obtained  from  Eq.~(\ref{a1a2}) through  Eqs.~(\ref{r})  and
~(\ref{vi}) taking  $\Delta{V}$=$\Delta{V}_{min}$ with $\Psi$=180$^{o}$, i.e.,
assuming the  impact in the  direction opposite to  the orbital motion  of the
satellites and taking the positive sign of the square root in Eq.~(\ref{r}):

\begin{eqnarray}
a_{1M}=\frac{G\; m_U \; (B'-A)2}{(\Delta {V}_{min})2 \;  (1-A)
(\sqrt{B'}- \sqrt{A})^{2}}\;\;,\;\;
a_{2M}=\frac{G\; m_U \; (B'-A)2}{(\Delta {V}_{min})2 \;  (1-B)
(\sqrt{B'}- \sqrt{A})^{2}}\;\;.
\label{emax}
\end{eqnarray}

For  each  A, we  calculate  the  value of  B  (B=  $B'/  ( 1+  m_{i}/m_{U})$)
corresponding  to the  transfer to  $a_{2M}$= $a$,  where $a$  is  the present
orbital semiaxis  of each one  of the Uranian  irregulars shown in  the second
column of  Table 1 in units of  $R_U$.  From Eq.~(\ref{e2m}), this  value of B
provides the minimum possible value of $e_{2min}$, $e_{2m}$, that the orbit of
each irregular  may acquire at impact  for each initial condition  A and every
initial condition for $T_0$, $\alpha$, $\Psi$ and $m_{i}$, i.e., if a transfer
of a given orbit (A,B) is not possible for $\Psi$=180$^{o}$, $T_0$= 20 hrs and
$\alpha=70^{o}$,  the  same transfer  (A,B)  is  not  possible for  any  other
incident  direction of  the impactor  and  for any  other value  of $T_0$  and
$\alpha$ either.

Since the orbits  of the irregulars are time  dependent, the orbital evolution
of the five Uranian irregulars known in 2002 was computed in PB02 by numerical
integration of the  equations of the elliptical restricted  three body problem
formed by  the Sun, Uranus and the  satellite.  In this paper,  we present the
orbital evolution of the nine  known Uranian irregulars for $10^{5}$ yrs using
the  symplectic integrator  of  Wisdom \&  Holman  (\cite{wisdom}), where  the
perturbations of the Sun, Jupiter,  Saturn and Neptune are included.  The mean
($e_{mean}$), maximum  ($e_{max}$) and minimum  ($e_{min}$) eccentricities are
shown in {\bf Table 2} for all the known Uranian irregulars.

The transfer of  each satellite from its original orbit to  the present one is
possible only for  those values of (A,B) which  satisfy the condition $e_{2m}$
$<$  $e_{max}$. A  satellite did  not exist  before the  impact if  it  has no
transfer and the satellites with the  widest range of transfers are those with
the highest probability of existing before the impact.

The  transfers  within a  range  of 20  $R_U$  around  each present  satellite
semiaxis ($a_{2M}$=  $a$$\pm$ 20 $R_U$;  $a$ taken from  Table 1) for  all the
Uranian irregulars are  shown in Fig.1.  There are  few transfers for Setebos,
Ferdinand and Margaret.   This makes the existence of  these satellites before
collision little probable.   The only transfers for Trinculo  and Prospero are
close to the pericenter of  an eccentric initial outer orbit ($e_{1m}$$>$ 0.58
for  Prospero and $e_{1m}$$>$  0.62 for  Trinculo).  The  minimum eccentricity
after collision $e_{2m}$ for Trinculo  is in the range [0.16-0.23], very close
to  $e_{max}$  (0.237). This  result  gives a  very  low  probability for  the
existence of  Trinculo before the GC.   For Prospero $e_{2m}$ is  in the range
[0.52-0.57],  $e_{2m}$  $\sim$ $e_{max}$  (0.571).   Therefore this  satellite
could not exist  before the GC.  If the present large  obliquity of Uranus was
caused  by a  large  impact  at the  end  of its  formation,  Prospero had  to
originate after the event.  Relating the origin of the outer Uranian system to
a  common  formation  process,   all  the  Uranian  irregulars  probably  were
originated  after the GC.   The possible  post-GC origin  of Prospero  and the
other Uranian irregulars is {\bf discussed} in the following section.

%----------------------------------------------------------------------------%

\begin{table} 
\centering                
\begin{tabular}{crrlclll}  
\hline \hline       
Satellite & $r_s[km]$  & $a[R_U]$ & $e_{mean}$ & $a_i[R_U]$ & $e_{i}$ &
 $(\Delta{a})/a_i)$ & $(\Delta{e}/e_i)$\\ \hline

Caliban&      49 &  283 & 0.191  & 287.5 & 0.1973 & 1.602 x 10$^{-2}$ &
 3.289 x 10$^{-2}$
\\ \hline

Sycorax &     95 &  482 & 0.541 & 485   & 0.5436 & 6.307 x 10$^{-3}$ &
 4.795 x 10$^{-3}$
\\ \hline

Prospero &     15 &  645 & 0.432  & 648   & 0.4342 & 4.585 x 10$^{-3}$  &
 4.997 x 10$^{-3}$
\\ \hline

Setebos &     15 &  694 & 0.581  & 701.8 & 0.5853 & 1.123 x 10$^{-2}$ &
 7.366 x 10$^{-3}$
\\ \hline

Stephano &     10 &  314 & 0.251  & 333   & 0.2781 & 6.058 x 10$^{-2}$ &
 0.1080
\\ \hline

Trinculo &     5  &  336 & 0.218  & 361.6 & 0.2501 & 7.623 x 10$^{-2}$ &
 0.1475
\\ \hline

Ferdinand &     6  &  813 & 0.660  & 839.3 & 0.6701 & 3.241 x 10$^{-2}$ &
 1.533x 10$^{-2}$
\\ \hline

Francisco &     6  &  169 & 0.142  & 280.6 & 0.3593 & 0.6604    & 1.530
\\ \hline

Margaret &  5.5& 579 & 0.633  & 649 & 0.6705  & 0.1209 & 5.917  x 10$^{-2}$ \\
\hline \end{tabular} \caption{Present parameters of the Uranian irregulars and
orbital damping  due to gas drag  exerted by Uranus  extended envelope.  $r_s$
and $a$  are the present physical  radius and the present  orbital semiaxis of
the irregulars.  $e_{mean}$ is their calculated mean eccentricity tabulated in
Table 2.   $a_i$ and  $e_{i}$ are the  orbital semiaxis and  eccentricity just
after the  GC, while $(\Delta{a})/a_i)$ and $(\Delta{e}/e_i)$  are the damping
of these orbital  elements since the epoch of the GC  until the contraction of
Uranus envelope.}  \label{damping} \end{table}

%---------------------------------------------------------------------------%

\begin{table} 
\centering                
\begin{tabular}{crrl}  
\hline \hline
Satellite &  $e_{mean}$   &  $e_{max}$   & $e_{min}$  \\ 
\hline
Caliban &  0.191  & 0.315  & 0.072  \\
\hline
Sycorax  &  0.514  & 0.594 &  0.438 \\
\hline
Prospero &   0.432 &  0.571 &   0.305  \\
\hline
Setebos &  0.581  &  0.704  &  0.463  \\
\hline
Stephano &  0.251 &  0.381 &  0.121  \\
\hline
Trinculo &  0.218  & 0.237   & 0.200  \\
\hline
Ferdinand &  0.660 &  0.970 &  0.393  \\
\hline
Francisco  &  0.142  & 0.187 &   0.093  \\
\hline
Margaret  &  0.633 &  0.854 &  0.430   \\
\hline
\end{tabular}
\caption{Variation of the eccentricity of the Uranian irregulars due
to Solar and giant planet perturbations over a period of $10^{5}$ yrs.}
\label{eccentricity} 
\end{table}

%----------------------------------------------------------------------------%

\begin{figure}       \centering      \includegraphics[width=16cm]{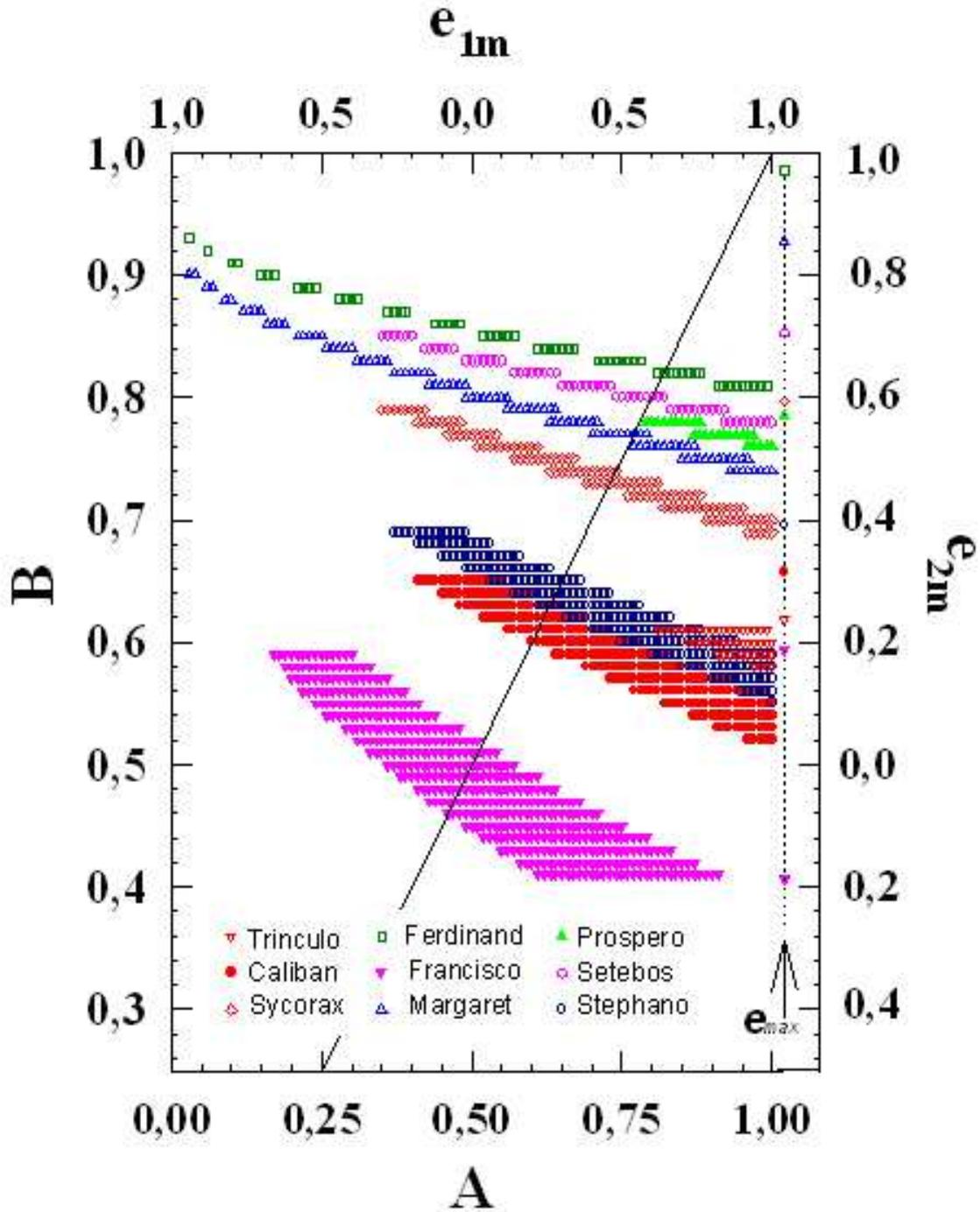}
\caption{The transfers capable of producing  the present orbits of the Uranian
Irregulars.  A  (B) is the square of  the ratio of the  satellite's speed just
before (after) the  impact to the escape velocity  at the satellite's location
just  before   (after)  the  impact.   $e_{1m}$  ($e_{2m}$)   is  the  minimum
eccentricity of the orbits before  (after) collision.  The full-black line A=B
divides the upper region (the current  orbits arise from inner orbits) and the
lower  region (the  current orbits  arise from  outer orbits).   The  value of
$e_{max}$ tabulated  in Table 2  is shown on  the dashed line  for comparation
with  $e_{2m}$.   Trinculo:  empty  down  triangles,  Caliban:  full  circles,
Sycorax:  empty  rhombus,  Ferdinand:  empty  squares,  Francisco:  full  down
triangles, Margaret: empty up triangles, Prospero: full up triangles, Setebos:
empty  hexagons,  Stephano:  empty  circles.}  \label{fig1} \end{figure}

%-----------------------------------------------------------------------------%

\section{Origin of Prospero after the Great Collision}

In this section,  we analyze the possibility that  Prospero have been captured
after  the GC.   We investigate  the possible  dissipative mechanisms  able to
produce its  permanent capture  taking into account  that the giant  impact is
assumed to have occurred at late stages in the planetary accretion process.

\subsection{Gas drag by Uranus' envelope}

Bodenheimer   \&    Pollack   (\cite{bodenheimer})   and    Pollack   et   al.
(\cite{pollacka}) studied the  formation of the giant planets  by accretion of
solids and gas. In their model,  the so called core instability scenario, when
the mass of the core of the  planet has grown enough a gaseous envelope begins
to  form around it.   For Uranus,  its envelope  extended until  its accretion
radius which was $\sim$ 500 $R_U$ at the end of Uranus' formation (Bodenheimer
\&  Pollack \cite{bodenheimer}).  The  formation of  Uranus is  completed when
there  is  no  more nebular  gas  to  accrete.   Otherwise, gas  accretion  by
proto-Uranus would have continued towards  the runaway gas accretion phase and
the planet would have now  a massive gaseous envelope.  Bodenheimer \& Pollack
(\cite{bodenheimer}) obtained that  after the end of accretion,  the radius of
the envelope of proto-Uranus remained  almost constant ($\sim$ 500 $R_U$) over
a time scale of 10$^{4}$ yrs and  then contracted rapidly to $\sim$ 8 $R_U$ in
10$^{5}$  yrs.  The  final  contraction to  the  present-day planetary  radius
occurred on a slower timescale of 10$^{8}$ yrs.

Korycansky et al.  (\cite{korycansky}) carried out hydrodynamical calculations
of the GC for a large set of initial conditions at the end of accretion.  They
found a  sharp transition between the cases  where almost all the  mass of the
envelope of  Uranus remained after  the impact and  those where it  was almost
entirely dispersed by the impact.  This implies that the impact should have not
dispersed the envelope,  as there would have been no  nebular gas to re-accrete
on  the planet.   They showed  that  the envelope  reacts hydrodynamically  at
impact and it expands outward. After the  shock the gas falls back on the core
over a timescale of few hours being the final result a readjustment instead of
a catastrophic transformation.  The  timescale for this hydrodynamical process
is much  shorter than  the orbital period  of the  irregulars which is  of the
order of years.   We may then assume  that the GC did not  change the envelope
density profile.  

The  extended envelope  of  Uranus could  be  in principle,  a  source of  gas
allowing  the capture  of  Prospero and  the  other irregulars  after the  GC.
Assuming that the GC did not change the envelope profile, we fit from Fig.1 of
Korycansky et al.  (\cite{korycansky})  the density profile of Uranus' gaseous
envelope before  the GC, $\rho_{g}=  c_{\mathrm{te}} R^{-4}$ g  cm$^{-3}$ with
$c_{\mathrm{te}}$= $10^{36}$ and $R$ being measured in cm.  It gives a nebular
density of  $\sim$ 4 $\times$  10$^{-13}$ g cm$^{-3}$  at the boundary  of 500
$R_U$ in agreement with the minimum mass nebula model.

As a first approximation, we compute the ratio of gas mass traversed by a body
of density $\rho_s$  and radius $r_{s}$ in a  characteristic orbital period P,
to the mass of the body.  Assuming  for the body a circular orbit of radius R,
we calculate the so-called $\beta$ parameter (Pollack et al. \cite{pollack}):

\begin{equation}
\beta= \frac{P}{\tau}= \frac{2 \pi R \rho_g \pi r_{s}^{2}}
{\frac{4}{3} \pi \rho_s r_{s}^{3}}
= \frac{3 \pi \rho_g R}{2 \rho_s r_{s}},
\label{beta}
\end{equation}

\noindent where $\tau$ is the characteristic timescale for changing any of the
orbital parameters. For the permanent  capture to occur $\beta$ cannot be very
small,  $\beta$   $\geq$  0.04   (Pollack  et  al.    \cite{pollack}).   Using
Eq.(\ref{beta}) and  assuming a  nebula density 10  times that of  the minimum
nebula model ($c_{\mathrm{te}}$= $10^{37}$) for an object the size of Prospero
($\rho_{s}$= 1.5  g cm$^{-3}$)  and at Prospero's  pericenter (R=  278 $R_U$),
$\beta$ $\sim$ 9 $\times$ 10$^{-5}$, which is too small to affect the orbit of
Prospero.

Following BP02, we now investigate with more detail the possible effect of gas
drag on the  Uranian irregulars after the GC due  to Uranus' extended envelope
before  its contraction  to its  present  state.  Following  the procedure  of
Adachi  et  al.  (\cite{adachi}),  we   obtain  the  time  variations  of  the
eccentricity $e$ and  semiaxis $a$ of each Uranian  irregular.  The drag force
per unit mass is expressed in the form:

\begin{equation}
F= - C  \rho_{g}  v_{rel}^{2},     C= \frac{C_{D}  \pi
r_{s}^{2}}{2m},
\label{drag2}
\end{equation}

\noindent  where $v_{rel}$  is the  relative  velocity of  the satellite  with
respect to  the gas.  In  computing the satellite  mass $m$, a  satellite mean
density  $\rho_{s}$  of  1.5 g  cm$^{-3}$  is  taken  for all  the  satellites
(http://ssd.jpl.nasa.gov/?sat\_ phys\_par).   The drag coefficient  $C_{D}$ is
$\sim$ 1  and $r_{s}$ is each  satellite radius taken for  each satellite from
Table 1.

Assuming that  the orbital elements  are constant within one  Keplerian period
(the  variations of $a$  and $e$  are very  small), we  consider the  rates of
change of the elements averaged over one period, that is:

\begin{eqnarray}
\mean{\frac{da}{dt}}= -\frac{C}{\pi}
(G (m_{i}+m_{U}) a)^{\frac{1}{2}}
\int_{0}^{2\pi} {\frac{ \rho_{g}(e^{2}+ 1+ 2 e
\cos{\theta})^{\frac{3}{2}}}
{(1+e \cos{\theta})^{2}} d\theta} \nonumber\\
\mean{\frac{de}{dt}}= -\frac{C}{\pi} (1-e^{2})
\left(\frac{G (m_{i}+m_{U})}{a}\right)^{\frac{1}{2}}
\int_{0}^{2\pi} {\frac{ \rho_{g} (e+ \cos{\theta})  (e^{2}+ 1+
2 e \cos{\theta})^{\frac{1}{2}}}
{(1+e \cos{\theta})^{2}} d\theta}.
\label{drag6}
\end{eqnarray}

Since after the end  of accretion the gas density in the  outer regions of the
envelope contracts rapidly, we  have integrated Eqs.(\ref{drag6}) back in time
on 10$^{4}$ yrs for the 9  Uranian irregulars.  We have taken $v_{rel}$ as the
satellite orbital  velocity since we have  assumed a null  gas velocity.  This
assumption  maximizes  the orbital  damping  for  retrograde satellites  which
allows  us  to  set  upper  bounds  in the  damping  effect  for  the  orbital
eccentricity  and  semiaxis  of  all  the  retrograde  irregulars.   The  mean
eccentricity $e_{mean}$ and the actual semiaxis $a$ from Table 1 were taken as
the   initial  conditions   for   the  integrations.    We  take   $\rho_{g}$=
$c_{\mathrm{te}}$ R$^{-4}$ g cm$^{-3}$  with $c_{\mathrm{te}}$ = 10$^{36}$ and
$R= a (1-e^{2})/ (1+ e \cos{\theta})$,  $a$ being measured in cm.  The orbital
damping is  shown in Table 1, where  $a_i$ and $e_i$ are  the initial semiaxis
and   eccentricity  just  after   the  GC   at  the   end  of   accretion  and
$\Delta{a}$=$(a_i-a)$  and $\Delta{e}$= $(e_i-e_{mean})$,  are the  damping in
the  orbital semiaxis and  eccentricity. Stephano,  Trinculo and  Margaret had
experienced  little orbital  evolution while  Francisco had  suffered  a large
orbital  damping  (see Table  1).   The  permitted  transfers of  Fig.1  would
increase for Francisco  since the condition $e_{2m}$ $<$  $e_i$ should be then
satisfied.   However,  the orbital  evolution  of  the  larger satellites,  in
particular  of  Prospero,  results  negligible.  Even  increasing  the  nebula
density by  a factor  of 10 ($c_{\mathrm{te}}$=10$^{37}$)  the damping  of the
orbital  elements of  Prospero is  too small  with $\Delta{a}/a_i$=  0.046 and
$\Delta{e}/e_i$=  0.048.   This  satellite  had not  experienced  any  orbital
evolution  due to  Uranus'  extended envelope  and  then could  not have  been
capture by gas drag after the GC.

The  pressure forces  acting on  a body  traveling through  the gas  not only
decelerates it,  but also subjects it  to stresses.  If the  stress is greater
than the strength of the body, the body is fractured in fragments of different
size. The  fragments move  away one another  since drag forces  vary inversely
with  size and  act to  separate them.   The average  pressure on  the forward
hemisphere of a non-rotating, spherical body  as it moves through the gas with
relative velocity  $v_{rel}$ is approximately  equal to the  dynamic pressure,
$p_{dyn}  = \frac{1}{2}  \rho_g v_{rel}^{2}$  (Pollack  et al.\cite{pollack}).
The  body will fragment  into pieces  if $p_{dyn}$  $\geq$ Q,  where Q  is the
compressive strength.   Values of Q on  the order of 3  $\times$ 10$^{6}$ dyne
cm$^{-2}$ are needed to shatter strong  (e.g.  rock/ice) targets ( which is 10
times lower than the value  adopted for asteroids), while compressive strength
on  the  order of  3  $\times$ 10$^{4}$  dyne  cm$^{-2}$  are appropriate  for
relatively  weak (snow-like)  targets (Farinella  \&  Davis \cite{farinellaa},
Stern \cite{stern}).  For  a body on circular orbit  at the present pericenter
of  Prospero (R=  278 $R_U$),  where $v_{rel}$  is the  circular  speed around
Uranus at  R and taking $c_{\mathrm{te}}$= 10$^{37}$  for $\rho_{g}$, $p_{dyn}
\sim$ 0.17 dyne  cm$^{-2}$, and at Sycorax's pericenter  $p_{dyn} \sim 1$ dyne
cm$^{-2}$.  In  both cases  $p_{dyn} <<$  Q and Prospero  could not  have been
originated by the dynamical rupture of a parent object.

A collision may  fracture the parent body  but if the energy at  impact is not
sufficient to  disperse the  fragments, drag forces  may act to  separate them
against their mutual attraction.  The  relative importance of these effects is
measured by the  ratio $\epsilon_{j}$ of the drag force on  a given fragment j
to the  gravitational force acting on j  by the other fragments  i (Pollack et
al.\cite{pollack}):

\begin{equation}
\epsilon_j=\sum_{i,i \neq j}{\frac {F_{Dj} m_j}{F_{Gij} m_j}},
\label{ej}
\end{equation}

\noindent where $F_{Gij}$  is the gravitational force between  the particles i
 and j, and $F_{Dj}$ is the drag force on the particle j of radius $r_j$:

\begin{equation}
F_{Dj}= \frac{C_D \rho_g \pi r_{j}^{2} v^{2}}{\frac{8}{3} \pi
 \rho_s r_{j}^{3}},
\label{fj}
\end{equation}

The fragment  j is dispersed  by the gas  if $\epsilon_{j}$ $>$ 1  (Pollack et
al.\cite{pollack}).   In order  to  estimate  the order  of  magnitude of  the
effect,  we computed  Eq.(\ref{ej})  for j=Prospero  at Prospero's  pericenter
taking into  account the gravitational  attraction due to another  fragment of
equal size,  and at Sycorax  pericenter taking into account  the gravitational
force of Sycorax on Prospero.  In the first case, we obtain $\epsilon_{j}$= 10
$^{-5}$ and in the second case at Sycorax pericenter $\epsilon_{j}$= 0.04.  We
then conclude that pressure forces are not strong enough neither to fragment a
possible parent  object from  which Prospero originated,  nor to  disperse its
fragments.  In addition Prospero suffered no orbital evolution due to gas drag
and could not have been capture by Uranus'envelope after the GC.

\subsection{Pull-down capture}

Within the  GC scenario, runaway of  the cores of the  planets occurred during
the first stages  of accretion but stopped for each embryo  after it reached a
size of  about 1000  Km.  At  10-35 AU the  final mass  distribution contained
several hundreds  of Mars-size (or larger)  bodies dominating the  mass of the
residual disk.   Beaug\'e et al. (\cite{beauge}), investigated  the effects of
the post-formation  planetary migration  on satellites orbits.   They obtained
that if the large-body component  (composed of Mars-size bodies) dominated the
mass  of the  residual disk,  the presently  accepted change  in the  orbit of
Uranus of $\sim$ 3 AU is too  large and it is not compatible with the observed
distribution  of its  satellites.  Even  an orbital  change of  $\sim$  1.5 AU
already causes  sufficient instabilities to eject all  the Uranian irregulars.
Pull-down capture caused by the orbital expansion of the planet could then not
be a plausible mechanism for the  origin of Prospero and the other irregulars.
Pull-down capture caused  by the mass growth of the planet  after the GC would
not be possible given the impact is assumed to have occurred at the end of the
accretion process when there was no more mass to be accreted by the planet.

\subsection{Collisionless interactions}

Within the framework of the restricted three-body problem, a capture is always
followed by an escape.  To end up  with a long term capture, the satellite has
to dissipate energy  in a short time.  The  entrance energy $\Delta{E}$ within
the gravitational field of the planet is (Tsui \cite{tsui}):

\begin{equation}
\Delta{E}= - 2.15 \mu^{2/3} (1-\delta)
\frac{G M_{\odot}}{a_p},
\label{tsui}
\end{equation}

\noindent $\mu$=  $M_p$/$M_{\odot}$ and $\delta  <<$ 1, where $M_p$  and $a_p$
are the mass and orbital semiaxis of the planet . 

Tsui  (\cite{tsui}), suggested  a permanent  capture mechanism  where  a guest
satellite  encounters some  existing inner  orbit massive  planetary satellite
causing its  velocity vector  to be deflected  keeping the irregular  in orbit
around  the planet. In  this way,  the effective  two-body potential  would be
about  twice the  entrance energy  $\Delta{E}$  of the  guest satellite.   The
radius $R_1$  of the  orbit of  the guest satellite  after deflection  is then
given by:

\begin{equation}
R_1= \frac{0.23}{1-\delta} \mu^{1/3} a_p.
\label{tsui1}
\end{equation}

In  the case  of Uranus,  for  a minimum  entrance energy  of $\delta$=0,  the
minimum permanent orbital  radius of the guest satellite  is $R_1$= 955 $R_U$.
This value of $R_1$ is much  larger than the present semiaxis of Prospero (see
Table 1), making the capture of Prospero by this mechanism implausible.

The fact that  binaries have recently been discovered in  nearly all the solar
system's  small-body  reservoirs  suggests  that  binary-planet  gravitational
encounters could bring  a possible mechanism for irregulars  capture (Agnor \&
Hamilton  \cite{agnor}).   One possible  outcome  of gravitational  encounters
between a binary system and a planet is an exchange reaction, where one member
of the  binary is expelled  and the other  remains bound to the  planet.  Tsui
(\cite{tsui})  extended  the   scenario  of  large  angle  satellite-satellite
scattering to the formation of the Pluto-Charon pair assuming that Pluto was a
satellite  of  Neptune  and  that  Charon  was  a  guest  satellite.   Through
Eqs.(\ref{tsui}) and (\ref{tsui1}), the conditions  for the escape of the pair
was found.   Following their scenario, let  us  consider  the hypothesis that
Prospero was  a member of a  guest binary entering Uranus'  field, with energy
density $\Delta{E}_{bin}$, above the  minimum density given by Eq.(\ref{tsui})
and $\delta$=0.  A  close encounter with Uranus could  result in disruption of
the binary,  leading to the ejection of  one member and capture  of the other.
The minimum semiaxis  $R_1$ is given by Eq.(\ref{tsui1}).   However, even this
scenario seems to  be unlikely since the semiaxis of  Prospero is smaller than
955 $R_U$.

\subsection{Collisional interactions: Break-up processes}

Collisional interactions between two  planetesimals passing near the planet or
between  a  planetesimal and  a  regular  satellite,  the so  called  break-up
process,  leads to  the  formation  of dynamical  groupings  (e.g. Colombo  \&
Franklin 1971, Nesvorny et  al. \cite{nesvornya}).  The resulting fragments of
each progenitor  body after  a break-up will  form a population  of irregulars
expected  to  have similar  surface  composition,  i.e.   similar colors,  and
irregular shapes, i.e.  large temporal  variations in the light curve as these
irregular bodies rotate.

The  critical  rotation period  ($T_{c}$)  at  which centripetal  acceleration
equals gravitational acceleration for a rotating spherical object is:

\begin{equation}
T_c= \bigg(\frac{3 \pi}{G \rho_{ob}}\bigg)^{1/2},
\label{eq:tc}
\end{equation}

\noindent where G is the gravitational constant and $\rho_{ob}$ is the density
of the  object. With $\rho_{ob}$=0.5,  1, 1.5 and  2 g cm$^{-3}$,  $T_c$= 4.7,
3.3, 2.7 and 2.3 hrs, respectively.   The rotation period of Prospero is about
4 hrs (Maris et al.  \cite{marisa}), and it seems unlikely that light and dark
surface markings  on a spherical Prospero  could be responsible  for its light
curve  amplitude of  0.2 mag  (Maris et  al. \cite{marisa}).   Even  at longer
periods,  real  bodies will  suffer  centripetal  deformation into  aspherical
shapes.  For a given density and  specific angular momentum (H), the nature of
the deformation depends  on the strength of the object.   In the limiting case
of a strengthless (fluid) body,  the equilibrium shapes have been well studied
(Chandrasekhar  \cite{chandrasekhar}).  For H  $\leq$ 0.304  [in units  of ($G
M^{3} R_{sphe})^{1/2}]$, where M (kg) is the mass of the object and $R_{sphe}$
is the radius of an equal-volume sphere, the equilibrium shapes are the oblate
Maclaurin spheroids.  For 0.304 $\leq$  H $\leq$ 0.390 the equilibrium figures
are triaxial  Jacobi ellipsoids.   Strengthless objects with  H $>$  0.390 are
rotationally  unstable to  fission.   The Kuiper  Belt  objects (KBOs),  being
composed  of solid  matter, clearly  cannot be  strengthless.  However,  it is
likely  that the  interior structures  of  these bodies  have been  repeatedly
fractured by impacts, and that their mechanical response to applied rotational
stress is  approximately fluid-like. Such  ``rubble pile'' structure  has long
been studied  in the asteroid  belt (Farinella et al.\cite{farinella})  and in
the  Kuiper  Disk (Sheppard  \&  Jewitt  \cite{sheppard},  Jewitt \&  Sheppard
\cite{jewitt},  Romanishin \& Tegler  \cite{romanishin}).  Farinella  \& Davis
(\cite{farinellaa}) obtained  that KBOs larger  than about 100 km  in diameter
are massive enough to survive collisional disruption over the age of the solar
system, but may nevertheless have been internally fractured into rubble piles.

Whether a  collision between  an impactor  and a target  results in  growth or
erosion  depends primarily  on  the energy  of  the impact  and  the mass  and
strength of the target.  If the mass  of the impactor is small compared to the
mass  of the  target  $m_s$, the  energy required  at  impact to  result in  a
break-up is given by:

\begin{equation}
\frac{1}{2} m_s v_{col}^{2} \geq m_s S + \frac{3}{5}
 \frac{G m_s^{2}}{\gamma R_{ms}},
\label{eq:break}
\end{equation}

\noindent where  $v_{col}$ is the collision  speed, S is  the impact strength,
$R_{ms}$  is the  radius  of the  target  and $\gamma$  is  a parameter  which
specifies the fraction  of collisional kinetic energy that  goes into fragment
kinetic  energy  and  is  estimated  to  be $\sim$  0.1  (Farinella  \&  Davis
\cite{farinellaa}).  The  speed of the  fragments is critical when  the target
has  a gravity  field. Fragments  moving slower  than the  local  escape speed
re-accumulate to form rubber pile structures. 

We attempt to investigate whether  Prospero could be a collisional fragment or
if a collision on a primary Prospero would result in a rubber pile structure.

In  computing  Eq.(\ref{eq:break}),  $v_{col}^{2}$ =  $v_e^{2}$+$v_{inf}^{2}$,
where $v_e$  is the  scape speed at  the target  surface and $v_{inf}$  is the
typical approach velocity of the two objects at a distance large compared with
the Hill sphere  of the target.  For two bodies colliding  in the Kuiper disk,
$v_{inf}$ is given by (Lissauer \& Stewart \cite{lissauer}):

\begin{equation}
v_{inf}^{2}= v_k^{2} \bigg(\frac{5}{4} e^{2}+i^{2}\bigg),
\label{eq:inf}
\end{equation}

\noindent  where $v_k$  is the  keplerian velocity,  $e$ is  the  mean orbital
eccentricity  and $i$  the  mean orbital  inclination  of the  KBOs.  We  take
$\mean{e}$=2$\mean{i}$ (Stern \cite{stern}).  Eq.(\ref{eq:break}) was computed
using Eq.(\ref{eq:inf})  for values of e  in the range  [0.01-0.1] and orbital
semiaxes of the  KBOs in the range  [30-60] AU.  In computing the  mass of the
target $m_s$,  we consider the radius $R_{ms}$  of the KBOs in  the range [10-
500] km and densities between [0.5  - 2] g cm$^{-3}$.  Impact strengths in the
range [3  $\times 10^{4}$-3  $\times 10^{6}$] erg  cm $^{-3}$ were  taken.  We
obtain that  targets with  $R_{ms}\leq$ 210 km  suffer disruption for  all the
values of these parameters in the  Kuiper Disk.  Prospero has a radius of just
only 15 km.   If Prospero originated from the Kuiper Belt,  it would have been
more likely a  collisional fragment rather than primary  body.  Prospero would
preserve an  irregular shape  after disruption since  it is such  small object
that is unable to turn  spherical because its gravity cannot overcome material
strength.  Prospero could have been  captured during the break-up event if the
two KBOs  collided within Uranus' Hill  sphere, which could be  possible for a
minimum orbital eccentricity of the original KBOs of 0.37.

We also consider the  case in which the target is a satellite of Uranus which
collides   with  a  KBO   that  enters   the  Hill   sphere  of   the  planet.
Eq.(\ref{eq:break}) remains valid but the following expression of $v_{inf}$ is
considered:

\begin{equation}
\vec{v}_{inf}= \vec{v}_{ip} - \vec{v}_{sp},
\label{eq:infsat}
\end{equation}

\noindent  where $\vec{v}_{ip}$ is  the velocity  of the  KB0 with  respect to
Uranus and $\vec{v}_{sp}$  is the satellite orbital velocity,  at the epoch of
th event.  We assume that the satellite orbital velocity is circular. In order
to  get bounds  in the  relative velocity,  we take  two values  of $v_{inf}$,
$v_{inf}$=$v_{ip} \pm v_{sp}$.  For $v_{ip}$, we assume that the KBO describes
a hyperbolic orbit around Uranus during the approach giving:

\begin{equation}
v_{inf}^{2}= \frac{2G(m_i+m_U)}{a_s}+ \frac{G M_{\odot}}{a_{kb}} ,
\label{eq:sat}
\end{equation}

\noindent  where we  assumed  $G M_{\odot}/a_{kb}$  as  the relative  velocity
between the KBO and Uranus far  from the encounter, $(m_i+m_U)$ is the present
mass of Uranus and $a_s$ the  orbital semiaxis of the satellite.  We calculate
Eq.(\ref{eq:break})   using  Eqs.(\ref{eq:infsat})   and   (\ref{eq:sat})  for
$a_{kb}$  in the range  [20,60] AU  and $a_s$  [100 -700]  $R_U$ for  the same
values of  S, $R_{ms}$  and densities  we have used  for the  collisions among
KBOs.  We obtain  that for all the possible  parameters, any Uranus' satellite
with radius $R_{ms} \leq$ 1000 km suffers disruption if it collides with a KBO
larger than 10  km.  This process would lead to the  formation of two clusters
of irregulars,  one associated to the  preexisting satellite and  the other to
the primary KBO.   This process has the disadvantage that  it is unlikely that
the preexisting satellite  were formed from a circumplanetary  disk as regular
satellites given the large orbital semiaxis required for this object.

Break-up processes predict orbital  clustering.  However, no obvious dynamical
groupings  are observed  at the  irregulars  of Uranus.   A further  intensive
search of more  faint irregulars around Uranus is needed in  order to look for
dynamical and physical families.

\subsection{The CG itself  as a possible capture mechanism}

We now  turn to the question  of whether the  GC itself could have  provided a
capture mechanism  (BP02).  Since all  the transfers with  $A > B'$ lead  to a
more  bound orbit, this  process might  transform a  temporary capture  into a
permanent one (see Section 2  and Fig.1).  Moreover, a permanent capture could
even occur from an heliocentric orbit (transfers with $A$=1).

It is interesting to estimate the number of objects $N$ in heliocentric orbits
at the  time of the  GC, at distances  from Uranus less  than or equal  to 300
$R_{U}$.  Assume  that the  GC occurred when  Uranus was almost  fully formed,
meaning   that  its   feeding  zone   was  already   depleted   of  primordial
planetesimals. We  assume that  the objects passing  near Uranus at  that time
were mainly escapees  from the Kuiper belt. Using the  impact rate onto Uranus
and  the  distributions  of  velocities  and diameters  given  by  Levison  et
al. (\cite{levison}), and assuming that  the mass in the transneptunian region
at the  end of the  Solar System  formation was 10  times its present  mass, a
back-of-the-envelope calculation gives  one object of diameter D  $\geq$ 20 km
passing at a distance R $\leq$ 300 $R_U$ from Uranus every 6 yrs at the end of
accretion (BP02).  The  typical crossing time $T_C$ among  protoplanets in the
outer Solar System is larger than  one millon years (Zhou et al. \cite{zhou}).
The number  of objects passing  near Uranus during  a timescale $T_C$  is then
167000, which gives  a probability of 6 $\times$ 10$^{-6}$  for the capture of
an object at  about 300 $R_U$ by  the GC.  This low rate  of incoming objects,
makes the possibility  of the capture of all  the irregulars from heliocentric
orbits difficult.   Even the capture of  a single object,  Prospero (note that
Prospero could not have an orbit bound to the planet before the GC), turns out
to be  low probable.  Since  temporary capture can  lengthen the time  which a
passing body can  spend near the planet, a more  plausible situation arises if
we  assume that the  GC could  produce the  permanent capture  of one  or more
parent objects which were orbiting temporarily around Uranus being the present
irregulars the result of a collisional break-up occurring after the GC.

    %----------------------------------------------------------------------------%

    \section{Discussion and Conclusions}

    It is usually believed that the large obliquity of Uranus is the result of
    a great tangential collision (GC)  with an Earth-sized proto-planet at the
    end of the accretion process.   We have calculated the transfer of angular
    momentum and  impulse at impact  and have shown  that the GC  had strongly
    affected the orbits  of Uranian satellites.  We calculate  the transfer of
    the  orbits of  the nine  known Uranian  irregulars by  the GC.   Very few
    transfers exist  for five of  the nine irregulars, making  their existence
    before the GC hardly expected.  In particular, Prospero could not exist at
    the time of  the GC.  Then, either Prospero had to  originate after the GC
    or the  GC did  not occur, in  which case  another theory able  to explain
    Uranus'  obliquity and  the formation  of the  Uranian  regular satellites
    would be  needed.  It is usually  believed that the  regular satellites of
    Uranus have accreted from material  placed into orbit by the GC (Stevenson
    et al. \cite{stevenson}).

    Within the  GC scenario,  several possible mechanisms  for the  capture of
    Prospero after the GC were investigated.  If the Uranian irregulars belong
    to individual captures and relating the origin of the outer uranian system
    to a common formation process,  gas drag by Uranus' envelope and pull-down
    capture seem to be implausible.  Three-body gravitational encounters might
    be  a source of  permanent capture.   However, we  found that  the minimum
    permanent  orbital radius of  a guest  satellite of  Uranus is  $\sim$ 955
    $R_U$ while the current semiaxis of  Prospero is 645 $R_U$.  The GC itself
    could provide a mechanism of permanent capture and the capture of Prospero
    could have occurred from a heliocentric orbit as is required within the GC
    scenario, but due to  the low rate of incoming objects it  turns out to be
    difficult.  Break-up  processes could be  the mechanism for the  origin of
    Prospero and  the other  irregulars in the  frame of  different scenarios.
    Prospero might  be a fragment  of a primary  KBO fractured by  a collision
    with another KBO.  The fragment could  have been captured by Uranus if the
    two KBOs had a minimum orbital  eccentricity of 0.37.  Prospero could be a
    secondary  member of  a  collisional family  originated  by the  collision
    between another satellite  of Uranus and a KBO  where the parent satellite
    of Prospero could have been captured  by any mechanism before or after the
    GC.   This process  has  the disadvantage  that  it is  unlikely that  the
    preexisting satellite  were formed from a circumplanetary  disk as regular
    satellites  given the  large orbital  semiaxis required  for  this object.
    Since  collisional  scenarios require  in  general  high collision  rates,
    perhaps the irregulars were originally much more numerous than now.  Then,
    Prospero  and also  the other  irregulars might  be the  result  of mutual
    collisions  among  hypothetical preexisting  irregulars  (Nesvorny et  al.
    \cite{nesvorny}, \cite{nesvornyb})  which could have been  captured by any
    other mechanism before the GC.

The  knowledge of  the  size and  shape  distribution  of irregulars  is
important to know their relation  to the precursor Kuiper Belt population.  It
could bring  valuable clues to  investigate if they are  collisional fragments
from break-up processes occuring at the Kuiper Belt and thus has nothing to do
with how they  were individually captured later by the planet,  or if they are
collisional fragments produced during or  after the capture event (Nesvorny et
al.  \cite{nesvorny},  \cite{nesvornyb}).  The differential  size distribution
of the  Uranian irregulars approximates a  power law with  an exponent $q$=1.8
(Sheppard et al.\cite{sheppardc}).  If we assume that the size distribution of
the nine  irregulars with  radii greater than  7 km  extends down to  radii of
about  1 km,  we  would expect  about 75  irregulars  of this  size or  larger
(Sheppard et al.\cite{sheppardc}).

The nuclei  of Jupiter  family comets are  widely considered to  be kilometer-
sized fragments produced collisionally in  the Kuiper Belt (Farinella \& Davis
\cite{farinellaa}).   Jewitt  et  al.   (\cite{jewittb})  compared  the  shape
distribution  of  cometary  nuclei  in  the  Jupiter  family  with  the  shape
distribution of  small main-belt asteroids of  similar size (1 km  -10 km) and
with  the  shape  distribution  of  fragments produced  in  laboratory  impact
experiments.   They  found that  while  the  asteroids  and laboratory  impact
fragments show  similar distribution of axis ratio  ($\mean{b/a}$ $\sim$ 0.7),
cometary nuclei  are more elongated  ($\mean{b/a}$ $\sim$ 0.6).   They predict
that if comets reflect their collisional origin in the Kuiper Belt followed by
sublimation-driven  mass loss  once inside  the orbit  of Jupiter,  small KBOs
should  have average shapes  consistent with  those of  collisionally produced
fragments (i.e.,$\mean{b/a}$  $\sim$ 0.7). To date, constraints  on the shapes
of only  the largest  KBOs are  available.  Prospero being  a bit  larger than
cometary nuclei,  displays a variability of 0.21  mag in the R  band (Maris et
al.  \cite{marisa}).   This corresponds  to an axis  ratio projected  into the
plane  of  the  sky,  b/a  of  0.8.   The knowledge  of  the  size  and  shape
distribution of irregulars would shed light in the size and shape distribution
of small KBOs as well as on the irregulars capture mechanism.

Colors are an  important diagnostic tool in attempting  to unveil the physical
status and  the origin of the  Uranian irregulars.  In particular  it would be
interesting  to  assess  whether  it  is  possible  to  define  subclasses  of
irregulars just looking  at colors, and comparing colors  of these bodies with
colors of minor bodies in  the outer Solar System.  Avalilable literature data
show a dispersion  in the published values larger than  quoted errors for each
Uranian irregular  (Maris et al.   \cite{marisa}, and references  therein). We
have concluded in Maris et al.  (\cite{marisa}), that the Uranian irregulars
are slightly red but they are not as red as the reddest KBOs. 

An  intensive  search  for  fainter  irregulars  and a  long  term  program  of
observations able to recover in  a self consistent manner light-curves, colors
and phase effects informations is mandatory.

%---------------------------------------------------------------------------%

\begin{acknowledgements} MGP research was  supported by Instituto Argentino de
      Radioastronom\'{\i}a,   IAR-CONICET,   Argentina   and   by   Centro   de
      Astrof\'{\i}sica,   Fondo   de   Investigaci\'on   avanzado   en   Areas
      Prioritarias, FONDAP number 15010003,  Chile.  MM acknowledges FONDAP for
      finantial support  during a visit to  Universidad de Chile.  Part of the
      work of MM has been  supported by INAF FFO-{\it{Fondo Ricerca Libera}} -
      2006.  AB      research     was      supported      by     IALP-CONICET.
      We appreciate the useful suggestions by the Reviewer, which have helped
      us to greatly improve this paper. 

      \end{acknowledgements}

%---------------------------------------------------------------------------%

%-------------------------------------------------------------------------%

\end{document}